\documentclass[letter]{jpsj2}

\title{%
Effect of Impurity Scattering on the
Nonlinear Microwave Response in 
High-$T_{\rm c}$ Superconductors} 

\author{%
Takanobu {\sc Jujo}
\thanks{E-mail address: jujo@ms.aist-nara.ac.jp}
}

\inst{%
Graduate School of Materials Science, Nara Institute of Science 
and Technology, Ikoma, Nara 630-0101
}

\recdate{\today}

\abst{%
We theoretically investigate intermodulation distortion 
in high-$T_{\rm c}$ superconductors. 
We study the effect of nonmagnetic impurities on the real 
and imaginary parts of nonlinear conductivity. 
The nonlinear conductivity is proportional to 
the inverse of temperature owing to the dependence of 
the damping effect on energy, which arises from the phase shift 
deviating from the unitary limit. 
It is shown that the final-states interaction makes 
the real part predominant over the imaginary part. 
These effects have not been included in 
previous theories based on 
the two-fluid model, enabling a consistent explanation for 
the experiments with the rf and dc fields. 
}

\kword{%
intermodulation distortion, impurity scattering, 
unconventional superconductor, nonlinear Meissner effect, 
microwave conductivity, vertex correction
}

\begin{document}
\maketitle

The high-temperature superconductors are attractive for 
use in microwave circuits because of their low surface resistance 
as compared to those of normal metals.~\cite{lancaster} 
This low-loss property is disturbed by the nonlinearity of 
its response to external fields. 
This nonlinearity means that the system is unstable 
with respect to changes in the input power. 
This effect makes superconductors unsuitable for practical applications. 
On the other hand, the nonlinear response is useful to investigate 
the intrinsic properties of superconductivity. 
It has been predicted that 
the nonlinear Meissner effect (NLME) shows a peculiar behavior 
in unconventional superconductors.~\cite{yip} 
This prediction is summarized in the following two points. 
One is that the nonlinear correction 
to the magnetic field penetration depth ($\lambda$) is proportional 
to the inverse of temperature ($T$). 
Then, the divergence at low temperatures yields 
a nonanalytic response. 
The other is that the nonlinear correction takes different 
values depending on the direction of the external field. 
These can be evidence of the existence of nodes 
in superconductors. 

Several experiments have been conducted on this effect; 
these experiments yield different results 
depending on the measurement methods. 
An experiment that measures the dependence of 
$\lambda$ on the magnetic field yielded a result that is inconsistent 
with the theoretical prediction.~\cite{bidinosti} 
(Neither the low-temperature upturn nor the angle dependence 
is observed.) 
Intermodulation distortion (IMD) is theoretically supposed to 
reflect the presence of the NLME.~\cite{dahm1}
An experiment on the IMD seemingly shows 
a result consistent with the theoretical prediction.~\cite{oates}
(Only the low-temperature upturn is observed. The angle dependence 
has not been investigated. In this sense, 
this experiment is inadequate to be considered as evidence for the NLME.) 

In this paper, we microscopically develop the theory of 
nonlinear microwave response and 
consider how this contradiction arises. 
In the IMD experiment, 
the power is measured, which is expressed as 
$P_{IMD}\propto |\Delta R_s+{\rm i}\Delta X_s|^2$~\cite{collado}. 
Here, $\Delta R_s=\Delta\sigma_1/(2\sigma_2)\sqrt{\omega\mu/\sigma_2}$ 
and $\Delta X_s=-\Delta\sigma_2/(2\sigma_2)\sqrt{\omega\mu/\sigma_2}$ 
for $\sigma_2>>\sigma_1$. 
($R_s$, $X_s$, and $\sigma=\sigma_1-{\rm i}\sigma_2$ are 
the surface resistance, surface reactance, 
and conductivity, respectively, 
and $\Delta$ implies the nonlinear correction.) 
Previous theories on IMD assume 
the validity of the two-fluid model 
in addition to that of the theory of Yip and Sauls; 
in these theories, 
only $\Delta\sigma_2$ is considered.~\cite{dahm2} 
(We show that this assumption does not necessarily hold.)  
This is sufficient when the response to 
the nonlinear dc field~\cite{bidinosti} is considered. 
In the case of the IMD, however, 
there is a contribution from $\Delta\sigma_1$ in general. 

In the linear response, it is known that 
$\sigma_2>>\sigma_1$ holds. 
On the other hand, the relationship between 
$\Delta\sigma_1$ and $\Delta\sigma_2$ is not known. 
Therefore, we calculate both real and imaginary parts 
of the nonlinear conductivity to determine which quantity is predominant. 
We have to specify a dissipation mechanism 
in order to estimate $\Delta\sigma_1$, 
though this is not the case for $\Delta\sigma_2$. 
The NLME comes into question 
at the low-temperature region, where the $1/T$-upturn 
is supposed to occur. 
Therefore, we mainly consider the effect of nonmagnetic impurities 
on the nonlinear microwave response. 
This is because with regard to dissipation, 
the impurity scattering effect is dominant 
at low temperatures and the electron-electron correlation 
is dominant near $T_{\rm c}$.~\cite{hirschfeld} 
In this sense, we do not consider the type of 
correlation effect 
that functions as the enhancement factor and can be effective 
in the response to a static external field.~\cite{jujo} 
The absence of the NLME under a nonlinear dc field can be 
explained by taking this effect into account. 
An explanation to the above contradictory behavior 
can be provided by 
combining this effect with the invalidity of the two-fluid model 
discussed here. 

\begin{figure}
\includegraphics[width=5cm]{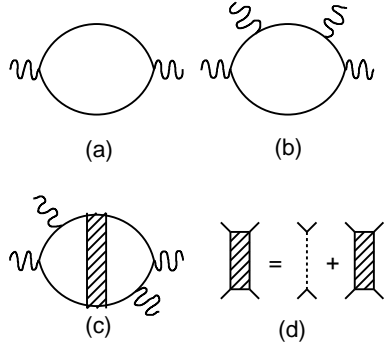}
\caption{(b) and (c) Representative diagrams for $K^{(3)}$. 
The solid and wavy lines denote 
the one-particle Green function and the external field, 
respectively. 
The shaded rectangles denote vertex correction. 
(a) Linear response diagram. 
(d) Diagram of the impurity scattering effect with 
the self-consistent t-matrix approximation.}
\label{fig:1}
\end{figure}

We consider isotropic impurity scattering. 
The self-energy with the self-consistent t-matrix approximation is 
\begin{equation}
\Sigma_0^R(\epsilon)=\frac{\Gamma_i G_0^R(\epsilon)}
{{\rm cot}^2\delta-G_0^{R}(\epsilon)^2}.
\end{equation}
Here, $\Gamma_i=n_i/\pi N(0)$ ($n_i$ and $N(0)$ are 
the impurity density and the density of states at the Fermi level 
in the normal state, respectively) 
and $G_0^R(\epsilon)={\rm Tr}\sum_k 
\hat{G}^R_{\epsilon,k}/(2\pi N(0))$ with 
the Green function 
\begin{equation}
\hat{G}^R_{\epsilon,k}=
\frac{1}{\tilde{\epsilon}^2-\xi_k^2-\Delta_k^2}
\begin{pmatrix}
\tilde{\epsilon}+\xi_k & \Delta_k \\
\Delta_k & \tilde{\epsilon}-\xi_k  
\end{pmatrix}. 
\end{equation}
($\tilde{\epsilon}=\epsilon-\Sigma_0^R(\epsilon)$.)
The nonlinear response function 
(third order) can be expressed as follows. 
(The vertex correction is given by the functional derivative 
of the self-energy by the one-particle Green function as that in the 
conserving approximation~\cite{baym}, which is also derived from 
Keldysh's method on the nonequilibrium state.~\cite{keldysh}) 
$K^{(3)}(\omega_1,\omega_2,\omega_3)=
\frac{1}{3!}\sum_{[i,j,k]}\int{\rm d}\epsilon
\tilde{K}^{(3)}_{\epsilon}(\omega_i,\omega_j,\omega_k)$. 
(The conductivity is expressed as 
$\Delta\sigma \propto K^{(3)}/\omega$.) 
$\sum_{[i,j,k]}$ means the sum of all permutations 
$\{i,j,k\}=\{1,2,3\}$ and $\omega=\omega_1+\omega_2+\omega_3$. 
\begin{equation}
\begin{split}
\tilde{K}^{(3)}_{\epsilon}(\omega_1,\omega_2,\omega_3)&=
{\rm Tr}[-f_{\epsilon4}
\hat{g}^{RRRR}_{\epsilon1\epsilon2\epsilon3\epsilon4}
-(f_{\epsilon3}-f_{\epsilon4})
\hat{g}^{RRRA}_{\epsilon1\epsilon2\epsilon3\epsilon4} 
-(f_{\epsilon2}-f_{\epsilon3})
\hat{g}^{RRAA}_{\epsilon1\epsilon2\epsilon3\epsilon4} \\
&-(f_{\epsilon1}-f_{\epsilon2})
\hat{g}^{RAAA}_{\epsilon1\epsilon2\epsilon3\epsilon4}
+f_{\epsilon1}
\hat{g}^{AAAA}_{\epsilon1\epsilon2\epsilon3\epsilon4}\\
+&{\rm Tr}[-f_{\epsilon4}
\{\hat{h}^{RRR}_{\epsilon4\epsilon1\epsilon2}
D^{RR}_{\epsilon2,\epsilon4}
\hat{h}^{RRR}_{\epsilon2\epsilon3\epsilon4}
+\hat{h}^{RRR}_{\epsilon3\epsilon4\epsilon1}
D^{RR}_{\epsilon1,\epsilon3}
\hat{h}^{RRR}_{\epsilon1\epsilon2\epsilon3} \} 
-(f_{\epsilon3}-f_{\epsilon4})
\hat{h}^{RAR}_{\epsilon3\epsilon4\epsilon1}
D^{RR}_{\epsilon1,\epsilon3}
\hat{h}^{RRR}_{\epsilon1\epsilon2\epsilon3} \\
&-(f_{\epsilon1}-f_{\epsilon2})
\hat{h}^{ARA}_{\epsilon4\epsilon1\epsilon2}
D^{AA}_{\epsilon2,\epsilon4}
\hat{h}^{AAA}_{\epsilon2\epsilon3\epsilon4} 
+f_{\epsilon1}
\{\hat{h}^{AAA}_{\epsilon4\epsilon1\epsilon2}
D^{AA}_{\epsilon2,\epsilon4}
\hat{h}^{AAA}_{\epsilon2\epsilon3\epsilon4}
+\hat{h}^{AAA}_{\epsilon3\epsilon4\epsilon1}
D^{AA}_{\epsilon1,\epsilon3}
\hat{h}^{AAA}_{\epsilon1\epsilon2\epsilon3}\}] \\
+&{\rm Tr}[-(f_{\epsilon3}-f_{\epsilon4})
\hat{h}^{ARR}_{\epsilon4\epsilon1\epsilon2}
D^{RA}_{\epsilon2,\epsilon4}
\hat{h}^{RRA}_{\epsilon2\epsilon3\epsilon4}
-(f_{\epsilon2}-f_{\epsilon3})
\hat{h}^{ARR}_{\epsilon4\epsilon1\epsilon2}
D^{RA}_{\epsilon2,\epsilon4}
\hat{h}^{RAA}_{\epsilon2\epsilon3\epsilon4} \\
&-(f_{\epsilon2}-f_{\epsilon3})
\hat{h}^{AAR}_{\epsilon3\epsilon4\epsilon1}
D^{RA}_{\epsilon1,\epsilon3}
\hat{h}^{RRA}_{\epsilon1\epsilon2\epsilon3}
-(f_{\epsilon1}-f_{\epsilon2})
\hat{h}^{AAR}_{\epsilon3\epsilon4\epsilon1}
D^{RA}_{\epsilon1,\epsilon3}
\hat{h}^{RAA}_{\epsilon1\epsilon2\epsilon3}]. 
\end{split}
\end{equation}
Here, 
$\hat{g}^{T1T2T3T4}_{\epsilon1\epsilon2\epsilon3\epsilon4}
=\sum_k v_k\hat{G}^{T1}_{\epsilon1,k}v_k\hat{G}^{T2}_{\epsilon2,k}
v_k\hat{G}^{T3}_{\epsilon3,k}v_k\hat{G}^{T4}_{\epsilon4,k}$, 
$\hat{h}^{T1T2T3}_{\epsilon1\epsilon2\epsilon3}
=\sum_k \hat{G}^{T1}_{\epsilon1,k}v_k\hat{G}^{T2}_{\epsilon2,k}
v_k\hat{G}^{T3}_{\epsilon3,k}$, 
$v_k$ is the quasiparticle velocity, 
$f_\epsilon={\rm tanh}(\epsilon/2T)$, 
$\epsilon_1=\epsilon$, $\epsilon_2=\epsilon-\omega_1$, 
$\epsilon_3=\epsilon-\omega_1-\omega_2$, $\epsilon_4=\epsilon-\omega$
, and $\omega=\omega_1+\omega_2+\omega_3$. 
$D^{RR}_{\epsilon,\epsilon'}$ and 
$D^{RA}_{\epsilon,\epsilon'}$ are vertex corrections; 
they are given afterward. 
The first and second traces represent the variations of 
the density of states and self-energy 
under the external field, respectively. 
The third trace implies the vertex correction, which 
represents the final-states interaction. 
The reasons for the invalidity of the application of the two-fluid model 
to the nonlinear response are as follows. 
(1) It is based on the 
assumption that the damping effect is independent of energy. 
(2) It includes only the nonlinear 
response of the density of states 
(the dependence of the damping effect on the external field 
and the final-states interaction are omitted). 
Therefore, we investigate these two aspects. 
The diagrams of the nonlinear response are shown in 
Fig.~\ref{fig:1}. 
Fig. 1(a) shows the diagram of the linear response; 
the nonlinear corrections are shown in (b) and (c). 
Fig. 1(b) and (c) show the nonlinear response with the variation of 
the density of states and vertex correction, respectively. 
In the linear response, 
vertex correction does not exist in the case of 
isotropic impurity scattering. 

First, we consider the nonlinear response 
arising from the variation of the density of states; 
its response function can be expressed as follows. 
\begin{equation}
\frac{{\rm Re}K^{(3)}_{DOS}}{\omega}\biggl|_{\omega\to 0}=
\int{\rm d}\epsilon \frac{\partial f_{\epsilon}}{\partial \epsilon}
\frac{\pi}{3}{\rm Re}\left(
\frac{\partial^2 n^{wR}_{\epsilon}}{\partial \epsilon^2}
\right)\frac{1}{\gamma_{\epsilon}}. 
\end{equation}
\begin{equation}
{\rm Im}K^{(3)}_{DOS}=
\int{\rm d}\epsilon \frac{\partial f_{\epsilon}}{\partial \epsilon}
\frac{2\pi}{3}{\rm Re}\left(
\frac{\partial^2 n^{wR}_{\epsilon}}{\partial \epsilon^2}
\right). 
\end{equation}
Here, $\gamma_{\epsilon}=-{\rm Im}\Sigma_0(\epsilon)$, 
$n^{xR}_{\epsilon}=\int_{\rm FS}V_x
\tilde{\epsilon}/
\sqrt{\tilde{\epsilon}^2-\Delta_k^2}$, and 
$V_{0,v,w}=1,v_k^2,v_k^4$. 
(We substitute $\Delta_k=\Delta_0{\rm cos}2\theta$ and 
take $\Delta_0$ as the unit of energy in the following 
numerical calculations.) 
If $\gamma_{\epsilon}$ is independent of energy, 
we have the same result as that when the two-fluid model is used. 

\begin{figure}
\includegraphics[width=8cm]{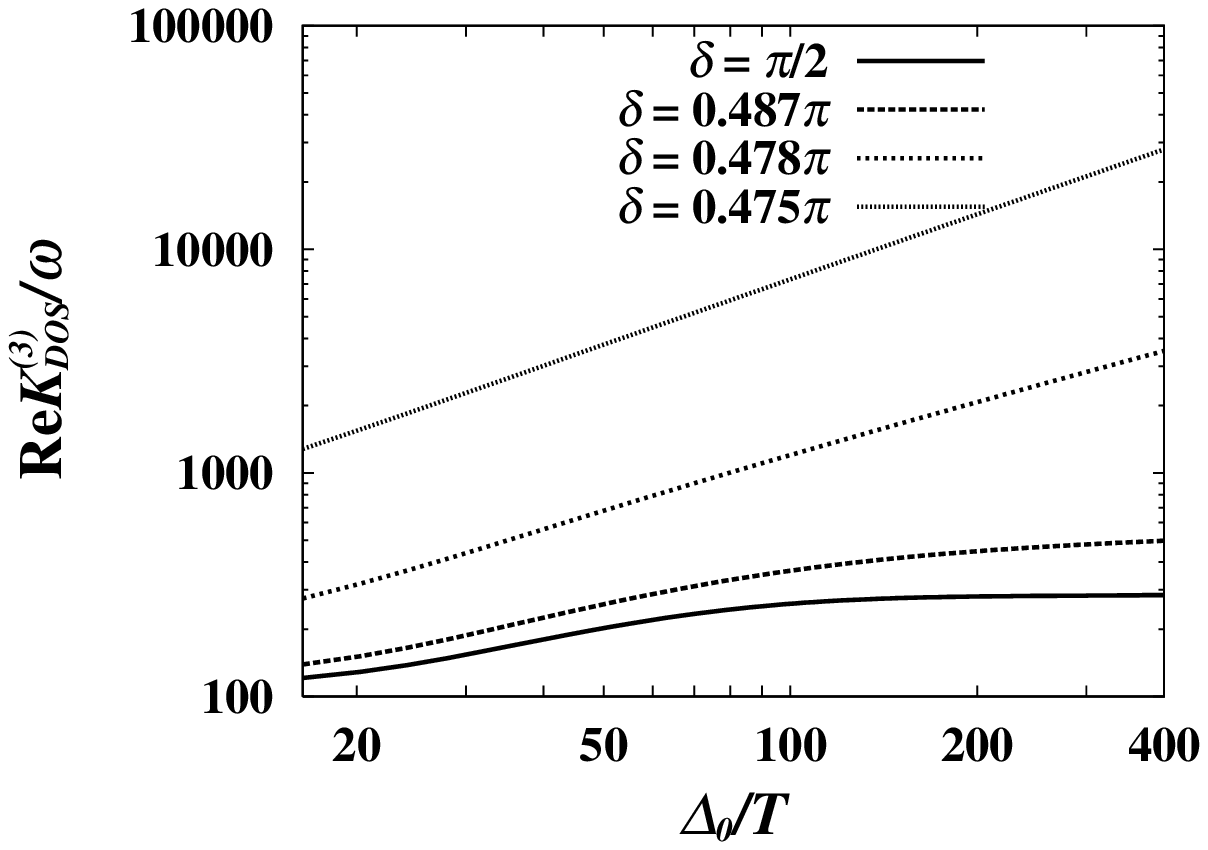}
\includegraphics[width=8cm]{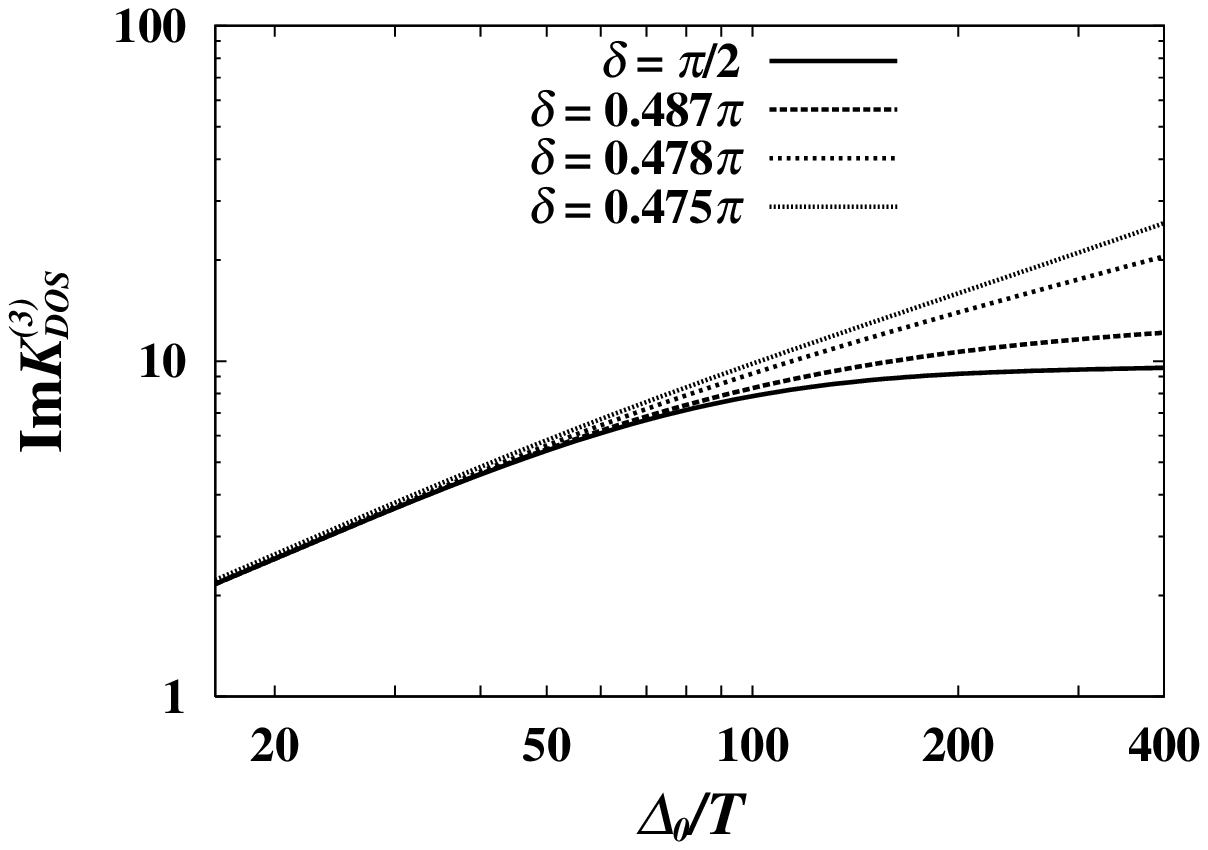}
\caption{Temperature dependences of 
${\rm Re}K^{(3)}_{DOS}/\omega|_{\omega\to 0}$ 
and ${\rm Im}K^{(3)}_{DOS}$ for various values of the phase shift 
$\delta$ and $\Gamma_i=0.001$.} 
\label{fig:2}
\end{figure}

\begin{figure}
\includegraphics[width=8cm]{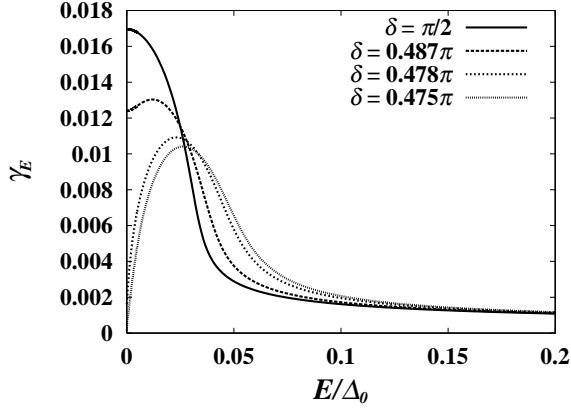}
\caption{Energy dependence of $\gamma_{\epsilon}$ 
with the same parameters as those used in Fig. 2.} 
\label{fig:3}
\end{figure}

The temperature dependences of ${\rm Re}K^{(3)}_{DOS}/\omega$ 
and ${\rm Im}K^{(3)}_{DOS}$ 
are shown in Fig.~\ref{fig:2}. 
As the phase shift $\delta$ deviates from unitary scattering 
($\delta=\pi/2$), 
${\rm Re}K^{(3)}_{DOS}/\omega|_{\omega\to 0}$ attains larger values 
and becomes proportional to the inverse of temperature. 
On the other hand, ${\rm Im}K^{(3)}_{DOS}$ does not show a clear 
$1/T$-divergence, but it is cut off at low temperatures. 
(The graph of $\delta=0.475\pi$ is seemingly divergent, 
but this is also verified to be cut off by comparing 
with that of smaller $\delta$ or $1/T$.) 
This behavior of $K^{(3)}_{DOS}$ can be explained 
by the dependence of the damping rate on energy. 
In previous theories on the nonlinear response in the Meissner 
state, the $1/T$-divergence is supposed to arise from 
the derivative of the density of states, which 
is cut off at low temperatures by the impurity scattering.~\cite{dahm3} 
(In clean systems, 
${\rm Re}\partial^2n_{\epsilon}/\partial\epsilon^2 \propto \delta(\epsilon)$ 
because ${\rm Re}n_{\epsilon}\propto |\epsilon|$.) 
If the damping rate takes a constant value, 
the result shown in Fig. 2 cannot be explained. 
The energy dependence of the damping rate $\gamma_{\epsilon}$ 
is shown in Fig.~\ref{fig:3}. 
As $\delta$ deviates from $\pi/2$, $\gamma_{\epsilon}$ 
decreases at around $\epsilon\simeq 0$. 
Then, 
${\rm Re}\partial^2 n^{wR}_{\epsilon}/\partial\epsilon^2$ 
increases, but it is cut off at low energies 
because $\gamma_0\ne 0$. 
Therefore, the different dependences on temperature arise in Fig. 2. 
${\rm Re}K^{(3)}_{DOS}/\omega$ shows a $1/T$-divergence 
owing to the energy-dependent damping effect. 
${\rm Im}K^{(3)}_{DOS}$ is cut off at low temperatures 
reflecting the energy dependence of  
${\rm Re}\partial^2 n^{wR}_{\epsilon}/\partial\epsilon^2$. 
 
The nonlinear correction to $K^{(3)}$ resulting from 
the variation of the self-energy is shown as a 
diagram similar to that in Fig. 1(c). 
The four-point vertex is expressed as 
$D^{RR}_{\epsilon,\epsilon}
=n_iT^{R2}_{\epsilon}/(
1-n_iT^{R2}_{\epsilon}
{\rm i}\pi\partial n^{0R}_{\epsilon}/\partial \epsilon)$. 
This term is small as compared to the vertex correction 
$D^{RA}_{\epsilon,\epsilon}$, 
which is verified by a numerical calculation. 
Therefore, we omit this term. 

Next, we consider the contribution of vertex correction to 
$K^{(3)}$, which is written as 
\begin{equation}
K^{(3)}_{VC}=
\int{\rm d}\epsilon \frac{\partial f_{\epsilon}}{\partial \epsilon}
\frac{\pi^2}{3}\sum_{[i,j,k]}\omega_k
(N_1+{\rm i}\omega N_2)
D^{RA}_{\omega-\omega_i}
(N_1+{\rm i}\omega_k N_2). 
\end{equation}
Here, 
$N_1={\rm Re}(\partial n^{vR}_{\epsilon}/
\partial \epsilon)/\gamma_{\epsilon}$, 
$N_2=[N_1/\gamma_{\epsilon}+{\rm Im}(\partial^2 n^{vR}_{\epsilon}/
\partial \epsilon^2)]/(2\gamma_{\epsilon})$, and 
$D^{RA}_{\omega-\omega_i}=
(\omega-\omega_i+2{\rm i}\gamma_{\epsilon})
/[\pi(\omega-\omega_i)
{\rm Re}n^{0R}_{\epsilon}/\gamma_{\epsilon}]$. 
(The term with $D^{RA}$ does not exist in the case of 
a nonlinear dc field.) 
The way in which the vertex correction depends on frequency 
originates from the identity 
\begin{equation}
\hat{\Sigma}^R_{\epsilon+\omega}-\hat{\Sigma}^A_{\epsilon}
=\Gamma_i\hat{T}^R_{\epsilon+\omega}\frac{1}{\pi N(0)}
\sum_k(\hat{G}^R_{k,\epsilon+\omega}-\hat{G}^A_{k,\epsilon})
\hat{T}^A_{\epsilon}, 
\end{equation} 
(here, $\hat{T}^R_{\epsilon}=(-{\rm cot}\delta\hat{\tau}_3
-\sum_k\hat{G}^R_{k,\epsilon}/\pi N(0))^{-1}$ and 
$\hat{\tau}_3=\begin{pmatrix}1 & 0 \\ 0 & -1 \end{pmatrix}$), 
which is similar to the identity discussed 
in the localization problem.~\cite{vollhardt} 

\begin{figure}
\includegraphics[width=8cm]{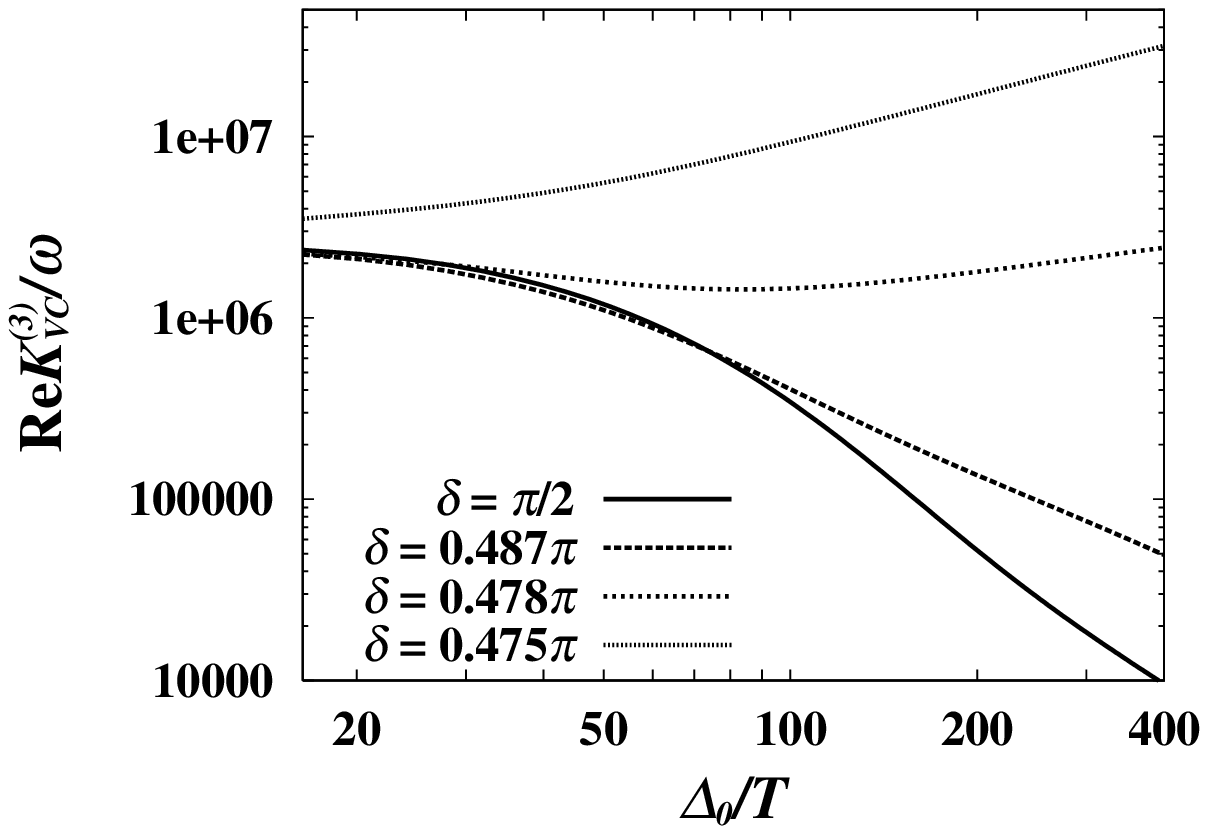}
\includegraphics[width=8cm]{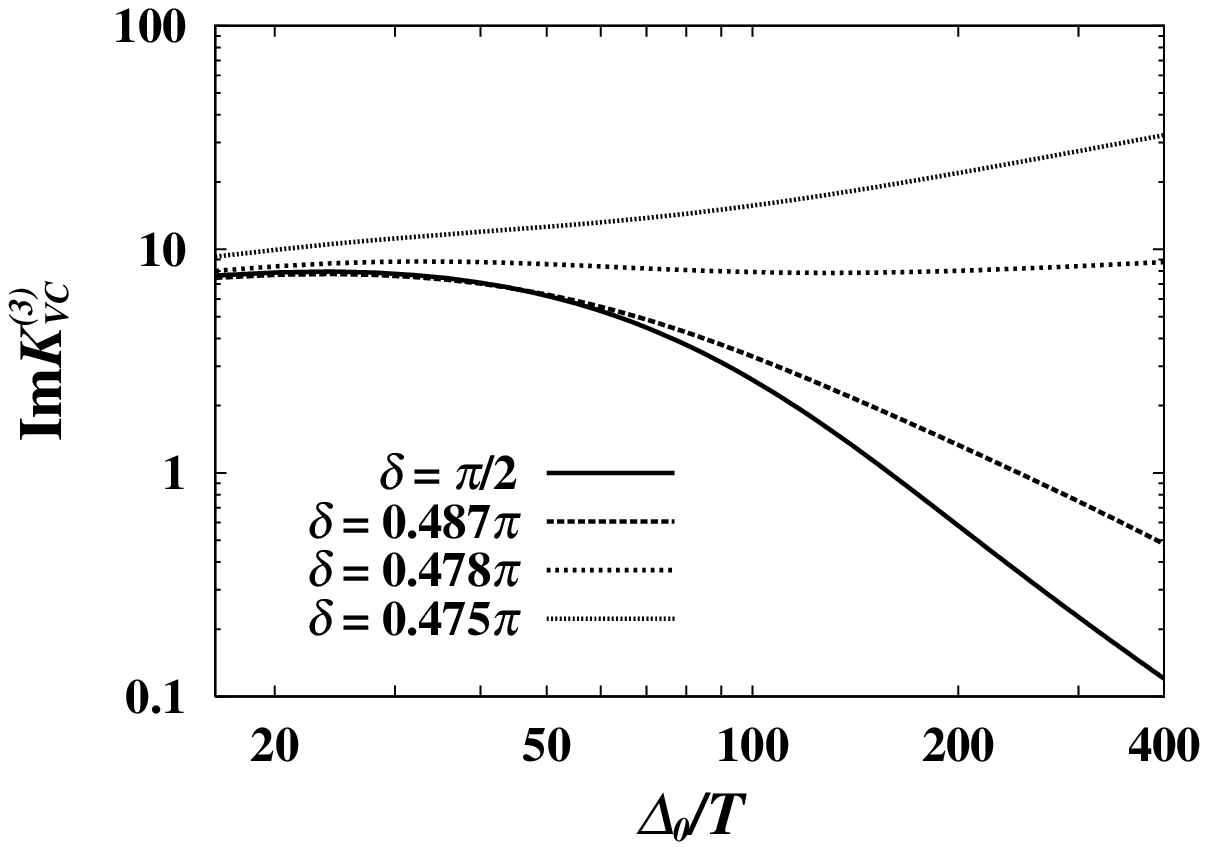}
\caption{Temperature dependences of ${\rm Re}K^{(3)}_{VC}/\omega$ 
and ${\rm Im}K^{(3)}_{VC}$ with various values of the phase shift 
$\delta$ and $\Gamma_i=0.001$. 
We substitute $\Delta\omega/\omega=0.01$.} 
\label{fig:4}
\end{figure}

In the numerical calculation of the two-tone IMD 
we substitute $\omega_1=\omega_2=\omega+\Delta\omega$ and 
$\omega_3=-\omega-2\Delta\omega$ and then 
maintain $\omega_{1,2,3}/\omega$ as constant for $\omega\to 0$. 
The contributions from the vertex correction, 
${\rm Re}K^{(3)}_{VC}/\omega|_{\omega\to 0}$ 
and ${\rm Im}K^{(3)}_{VC}$, 
are shown in Fig.~\ref{fig:4}. 
At $\delta=\pi/2$, 
both ${\rm Re}K^{(3)}_{VC}/\omega|_{\omega\to 0}$ 
and ${\rm Im}K^{(3)}_{VC}$ decrease as the temperature decreases. 
As the phase shift deviates from $\pi/2$, they show an upturn 
as $1/T$ increases. These behaviors are explained by 
the energy dependence of the damping rate and its effect on 
the density of states. 
Both ${\rm Re}K^{(3)}_{VC}/\omega|_{\omega\to 0}$ 
and ${\rm Im}K^{(3)}_{VC}$ are independent 
of the phase shifts at high temperatures. 
This means that the impurity scattering effect is less 
dependent on the phase shifts in this temperature region, 
as shown in the high-energy part of Fig. 3. 
The dependences of $K^{(3)}$ on phase shifts 
appear in the low-temperature region. 
The expression of $K^{(3)}$ indicates that 
${\rm Re}K^{(3)}_{VC}/\omega|_{\omega\to 0}$ 
and ${\rm Im}K^{(3)}_{VC}$ are proportional to 
$\gamma_{\epsilon}^{-1}$ and $\gamma_{\epsilon}^{0}$, 
respectively. 
This distinguishes the behaviors of the real 
and imaginary parts of $K^{(3)}$. 
${\rm Re}K^{(3)}_{VC}/\omega|_{\omega\to 0}$ shows 
an almost $1/T$-divergence, but 
${\rm Im}K^{(3)}_{VC}$ is roughly proportional to $1/\sqrt{T}$. 
The absence of the cut off at low temperatures 
in ${\rm Im}K^{(3)}_{VC}$ (unlike the case of ${\rm Im}K^{(3)}_{DOS}$) 
originates from the energy dependence of the density of states 
($n^{0R}_{\epsilon}$) in the vertex correction $D^{RA}$. 

\begin{figure}
\includegraphics[width=8cm]{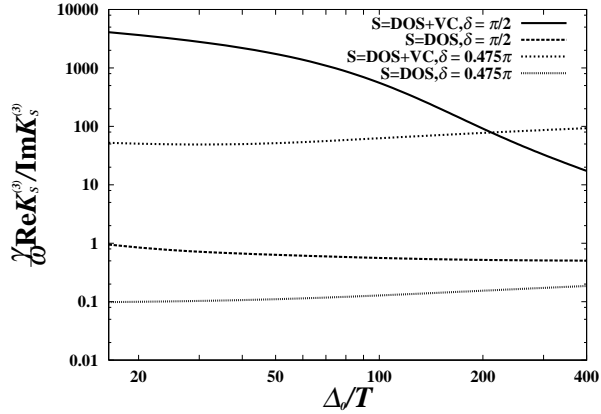}
\caption{Temperature dependences of 
$\gamma_0({\rm Re}K^{(3)}_{S}/\omega)/{\rm Im}K^{(3)}_{S}$ 
(here, $S=DOS$ or $DOS+VC$) 
with various values of phase shifts 
and the same parameters as those used in Fig. 4.} 
\label{fig:5}
\end{figure}

In Figs. 2 and 4, we can see that the real part of $K^{(3)}$ 
shows a $1/T$-divergence at some values of phase shifts. 
On the other hand, ${\rm Im}K^{(3)}$ does not show such a behavior. 
We should clarify 
which of $\Delta\sigma_1$ and $\Delta\sigma_2$ is predominant 
in order to specify the origin of the low-temperature upturn 
in the IMD power. 
To see this, we evaluate the following ratio. 
$\gamma_0({\rm Re}K^{(3)}/\omega)/{\rm Im}K^{(3)}$, 
which is equivalent to 
$(\gamma_0/\omega)\Delta\sigma_1/\Delta\sigma_2$, 
is shown in Fig.~\ref{fig:5}. 
In the hydrodynamic regime, which is the premise of our calculation, 
$\gamma_0$ is greater than $\omega$. 
Therefore, $\Delta\sigma_2$ is always predominant over 
$\Delta\sigma_1$ 
if $(\gamma_0/\omega)\Delta\sigma_1/\Delta\sigma_2<1$ holds. 
On the other hand, there is a possibility of 
$\Delta\sigma_1>\Delta\sigma_2$ 
in the case of $(\gamma_0/\omega)\Delta\sigma_1/\Delta\sigma_2>1$, 
depending on the value of $\gamma_0/\omega$. 
As shown in Fig. 5, 
if we consider only $K^{(3)}_{DOS}$, 
$\Delta\sigma_2>\Delta\sigma_1$ 
holds in the same way as the two-fluid model. 
When we take account of $K^{(3)}_{VC}$, 
$\Delta\sigma_1$ can predominate $\Delta\sigma_2$. 
As can be seen from Figs. 2 and 4, ${\rm Im}K^{(3)}_{VC}$ 
takes values of the same order as ${\rm Im}K^{(3)}_{DOS}$. 
On the other hand, ${\rm Re}K^{(3)}_{VC}/\omega$ takes values 
that are 100 times greater than ${\rm Re}K^{(3)}_{DOS}/\omega$. 
This difference originates from the following fact. 
It can be shown that the term $D^{RA}\propto 1/\Delta\omega$ 
arises in the real part of $K^{(3)}_{VC}$ 
(this term is cut off by the nonlocal effect mentioned below), 
but this term is canceled out in the imaginary part. 
Therefore, $\Delta\sigma_1$ can possibly predominate 
$\Delta\sigma_2$; as a result, the $1/T$-divergence 
can be originated from $\Delta\sigma_1$. 
This yields a solution for the contradiction 
between the experiments with the nonlinear rf and 
dc fields, which is not resolved when the two-fluid model is used. 

Here, we mention some issues that are not discussed above. 
Strictly speaking, in the case of $\delta\ne \pi/2$, the self-energy 
should be written as the matrix 
$\hat{\Sigma}^R(\epsilon)=
\Sigma^R_0(\epsilon)\hat{\tau}_0
+\Sigma^R_3(\epsilon)\hat{\tau}_3$. 
($\Sigma_3^R(\epsilon)=-\Gamma_i {\rm cot}\delta/[
{\rm cot}^2\delta-G_0^{R}(\epsilon)^2]$ and 
$\hat{\tau}_0$ is a unit matrix.) 
In this paper, we present the formula 
with $\Sigma_3^R(\epsilon)\to 0$ because 
it yields an intricate expression of $K^{(3)}$, 
and this gives almost the same numerical results 
as those for $\Sigma_3^R(\epsilon)\ne 0$. 
We present the numerical results of $K^{(3)}$ calculated by using the 
original expressions ($\Sigma_3^R(\epsilon)\ne 0$). 
The diamagnetic terms that include the factor 
$\partial v_k/\partial k\hat{\tau}_3$ are omitted. 
This is because out of the two branches, 
only the gap-full branch remains 
in the vertex correction $\hat{D}^{RA}$; 
these branches  arise from the matrix structure 
in the superconducting state. 
The nonlocal effect is not considered here. 
This effect also broadens the singular behavior 
of the derivative of the density of states in the same way 
as that by the impurity scattering effect.~\cite{jujo,kalenkov} 
However, the thickness of 
the film used in the IMD experiments~\cite{oates} is 
nearly 4000 \AA, which is 
almost 100 times thinner than that of the experiment with 
the nonlinear dc field~\cite{bidinosti}. 
This is almost the same order of magnitude as $\lambda$. 
Therefore, we omitted this effect here. 
(The numerical calculation of 
the current distribution with various 
values of $\lambda$ is 
given in ref. 15.) 
We show only the numerical results in which 
the impurity concentration $\Gamma_i$ was fixed. 
This is because our argument on $K^{(3)}$ can be 
similarly discussed when $\Gamma_i$ is varied. 
The different points are that 
the phase shift at which the $1/T$-divergence appears 
depends on $\Gamma_i$ and the absolute value of 
$K^{(3)}$ varies with $\Gamma_i$. 

In our theory, whether $P_{IMD}\propto1/T^2$ or not 
depends on the value of the phase shift, 
which is not known so far. 
As for the phase shift deviating from the unitary limit, however, 
there are several discussions related to the low-temperature 
thermal conductivity that suggests 
neither unitary nor Born limits.~\cite{kim,lofwander}
With regard to the comparison between the real and imaginary 
parts of the nonlinear conductivity, 
there is an experimental suggestion 
that $\Delta R_s$ is predominant over $\Delta X_s$~\cite{hott}, 
though $1/T$-divergence is not expected to exist 
in their temperature range. 
One of the possible experiments that can verify our theory 
is the third harmonic generation. 
When $\omega_1=\omega_2=\omega_3$, 
the contribution from vertex correction to ${\rm Re}K^{(3)}$
is reduced to the same order as ${\rm Re}K^{(3)}_{DOS}$. 
Therefore, it is expected that $\sigma_2>\sigma_1$ holds and 
the $1/T$-divergence is cut off at low temperatures. 

In this paper, we derived the general formalism of the nonlinear 
microwave conductivity under the influence of nonmagnetic impurities. 
We evaluated this formula by varying the value of 
the impurity scattering phase shift. 
As the phase shift deviates from the unitary limit, 
the nonlinear response shows a $1/T$-divergence 
owing to the dependence of the damping rate on energy. 
This is one of differences from previous theories 
where the $1/T$-divergence originates from 
the second derivative of the density of states. 
The predominance of the resistive part over 
the reactive part arises when the vertex correction is included. 
This term is not included in the two-fluid model. 
Therefore, the upturn of the IMD power at low temperatures 
can originate in the resistive part. 
This upturn does not need to be accompanied with 
$1/T$-divergence in the reactive part; 
this is a possible explanation to the seemingly 
contradictory results 
between the static and microwave experiments. 

The numerical computation in this study was carried out at 
the Yukawa Institute Computer Facility.

\end{document}